\documentclass[aps,showpacs]{revtex4}
\usepackage{epsf,latexsym}

\begin{document}

\title{From Unruh temperature to generalized Bousso bound}
\author{Alessandro Pesci}
\email{pesci@bo.infn.it}
\affiliation
{INFN, Sezione di Bologna, Via Irnerio 46, I-40126 Bologna, Italy}

\begin{abstract}
In a classical spacetime satisfying Einstein's equation
and the null convergence condition,
the same quantum mechanical effects
that cause black holes to have a temperature
are found to imply, 
if joined to the macroscopic nature of entropy,
the covariant entropy bound in its generalized form. 
This is obtained from
thermodynamics, 
as applied across
the local Rindler causal horizon
through every point $p$
of the null hypersurfaces $L$ the covariant entropy bound refers to,
in the direction of the null geodesics generating $L$.
\end{abstract}


\maketitle

$ $

After the works of Bekenstein \cite{Bekenstein} and
the theoretical discovery of quantum Hawking radiation \cite{Hawking},
the laws of black hole classical mechanics \cite{Bardeen}
turned to be the laws of thermodynamics as applied to black holes,
with $\kappa /2\pi$
being the physical temperature
and $A/4$ the physical entropy
of the black hole in fundamental units
($\kappa$ is the surface gravity and $A$ the horizon area of
the black hole).
Accordingly, a generalized second law of thermodynamics (GSL) has been introduced
(formulated by Bekenstein \cite{Bekenstein} prior to the discovery of Hawking radiation)
where the generalization is in that, in the computation of entropy,
in addition to ordinary matter entropy,
black hole entropy must be considered too.

This area-scaling black hole entropy has been suggested to be the maximum allowed entropy
for a system bounded by the given area $A$ \cite{tHooft1, Bekenstein94, tHooft2, Susskind}

\begin{eqnarray}\label{EntropyBound}
S \leq A/4.
\end{eqnarray}  
This is related
to the so called holographic principle, 
i.e. to the hypotesis that the physics of any spatial region
can be fully described by the degrees of freedom living on the boundary of
that region \cite{tHooft2, Susskind}.
Bousso covariant entropy bound \cite{Bousso}
appears to be a covariant reformulation
of bound (\ref{EntropyBound}) to give it a fully general validity and
overcoming some inadequacies
(and improving a previous suggestion of Fischler and Susskind \cite{Fischler}).
It can be stated as follows: 
for a spacetime with Einstein equation and satisfying certain energy conditions,
if $A$ is the area of any connected 
spacelike 2-surface $B$
and $L$ is a null hypersurface generated by surface-orthogonal null geodesics
with non-positive expansion, then

\begin{eqnarray}\label{BoussoBound}
S(L) \leq A/4,
\end{eqnarray}
where $S(L)$ is the entropy on $L$.

In \cite{FlanaganMarolfWald, BoussoFlanaganMarolf, Strominger}
some sufficient conditions for the validity of Bousso bound are given
in its generalized form (introduced in \cite{FlanaganMarolfWald}).
This generalized version states that if the geodesics generating $L$
are allowed to terminate at some spacelike 2-surface with area $A^\prime$
before reaching a caustic or a singularity, then (\ref{BoussoBound})
can be replaced by 

\begin{eqnarray}\label{GenBoussoBound}
S(L) \leq \frac{1}{4} (A-A^\prime)
\end{eqnarray}
and clearly it implies bound (\ref{BoussoBound})
as a particular case. 
Very interestingly, if specialized to the event horizons of black holes 
the bound (\ref{GenBoussoBound}) 
implies also the GSL (in its `classical limit' formulation,
that is not including the entropy of Hawking radiation \cite{Strominger}), 
provided we assume the validity
of the ordinary second law \cite{FlanaganMarolfWald}. 
It could then have the status of generalization
of the GSL to cover the case of arbitrary horizons \cite{Wald00}.
If so, the same physical premises bringing to state that
$A/4$ is the entropy of black hole horizon 
and as a consequence to the GSL, could
as well bring to the bound (\ref{GenBoussoBound}). 

The possibility to assign and
compute an entropy for black holes
arises at the end from the discovery that black holes have a temperature
of quantum origin \cite{Hawking, Gibbons}.
This temperature in turn, finds a local description
in terms of the Unruh temperature 
seen by accelerated observers in flat spacetime \cite{Unruh}.
Unruh temperature then reflects and requires the fundamental physical premise 
(the quantum principle) for black hole entropy; 
the mere introduction of this temperature could then imply, from what we said,
the generalized Bousso bound.
The aim of this Letter is precisely to investigate if this is the case.

Let us start from
a work of Jacobson \cite{Jacobson}.
From thermodynamics applied
at local Rindler horizons
attached to every point $p$ of spacetime
(local horizons characterized by vanishing expansion and shear at $p$,
see \cite{Jacobson}),
it is shown that if the horizon entropy
is assumed to be proportional to the area,
then
Einstein equation is equivalent to Clausius relation
$\delta Q^\prime = T^\prime \ dS^\prime$
(and the constant of proportionality above is 1/4)
where 
$dS^\prime$ 
is the variation of horizon entropy
(or the variation of the entropy 
of the system beyond the horizon in the inaccessible region,
as perceived by the accelerated observer) and 
$\delta Q^\prime$ and $T^\prime$ 
are respectively the energy flux 
(negative when energy is
leaving the system beyond the horizon)
and the Unruh temperature
as measured by an observer
just inside the horizon.
Reversing the logic, if we assume that both Einstein
and Clausius equations hold, we are forced to the following relation

\begin{eqnarray}\label{prescription}
dS^\prime = {\delta A}/4
\end{eqnarray}
between the variation of horizon entropy
and the variation $\delta A$ of the cross-sectional area $A$ 
of a pencil of generators of the local
Rindler horizon, 
determined by the focusing due to the energy flux
(just the amount needed for Einstein equation to hold). 
This prescription is not surprising as it is consistent
with the well known formula for the entropy 
of a Rindler horizon in flat spacetime \cite{Laflamme},
as well with what expected for arbitrary horizons
(see \cite{Padmanabhan,Amsel}).

Considering the 2-surface $B$ introduced above
with the congruence of surface-orthogonal null geodesics, 
what we need however is to put in relation the focusing acquired by the geodesics
while generating $L$ when they encounter some matter energy element 
with the entropy content $dS$ of this latter and not with $dS^\prime$.
The fact that these two entropies will in general differ
is consistent with the statement that entropy is in a sense an
observer-dependent concept
(as long as different are the degrees of
freedom which are hidden to different observers) \cite{Padmanabhan}.

Assuming to be in conditions that matter entropy can be described
by a 4-vector entropy flux $s^a$ \cite{FlanaganMarolfWald},
we proceed then to relate the above variation $\delta A$
of the cross-sectional area of the generators of
the future local Rindler horizon in the direction of the geodesics
generating $L$
in some small affine interval $\Delta\lambda = \eta$
with the entropy $dS = \int_{dL} s^a \epsilon_{abcd}$
spanned by the horizon generators in the same interval
(here $dL$ is the null hypersurface spanned, 
$\epsilon_{abcd}$ is the Levi-Civita tensor and
the notation of Appendix B of \cite{Wald}
is used for integrals of differential forms).
Under local thermodynamic equilibrium conditions,
from Raychaudhuri equation with vanishing expansion and shear
(and vanishing twist, being the horizon generators 
obviously surface orthogonal on the horizon)
we have

\begin{eqnarray}
\nonumber
- \frac{\delta A}{4} = \frac{1}{4} \int_{0}^{\eta} \left( 
\int_{0}^{\lambda} R_{ab} k^a k^b d{\lambda^\prime} \right) d\lambda \ A =
\frac{1}{4} \int_{0}^{\eta} \lambda R_{ab} k^a k^b d\lambda \ A =
\end{eqnarray}

\begin{eqnarray}\label{relate_1}
= 2\pi \int_{0}^{\eta} \lambda T_{ab} k^a k^b d\lambda \ A =
2\pi \int_{0}^{q \eta} \frac{l}{q} (\rho + p)
q^2 \frac{1}{q} d{l} A = \pi (q \eta)^2 (\rho + p) A =
\pi q \eta T \ \left[ \frac{\rho + p} T (q \eta) A \right], 
\end{eqnarray}
where use of Einstein equation has been made and the components of
stress-energy tensor $T_{ab}$ have been explicitly reported,
for the case of a perfect fluid.
Here $k^a$ is the null vector $k^a = (d/d\lambda)^a$,
$l \equiv q \lambda$ is the proper length in the fluid rest frame,
$R_{ab}$ is the Ricci tensor,
and $\rho$, $p$ and $T$ are local energy density, pressure and temperature
respectively.
The term in square brackets in the last expression is the entropy
spanned by the pencil of generators
with cross-sectional area $A$
in the fluid proper time $q \eta$
when the fluid,
supposed for simplicity homogeneous, 
has vanishing chemical potential $\mu$;
in the general case with $\mu \geq 0$, 
using the Gibbs-Duhem relation $\rho = T s - p +\mu n$
($s$ is here local entropy density),
we have $s \leq \frac{\rho + p}{T}$ so that,
if the null convergence condition
is satisfied
($T_{ab} k^a k^b \geq 0$ for all null vectors $k^a$),
equation (\ref{relate_1}) gives

\begin{eqnarray}\label{key}
dS \leq 
- \frac{1}{\pi (q \eta) T} \ \frac{\delta A}{4}.
\end{eqnarray}
This is our key equation as it relates 
the focusing acquired by the geodesics
when they encounter some matter ($\delta A < 0$) 
with its entropy content.
Even if shown only for the case of a perfect fluid,
this relation could be expected to be valid for general fluids  
at thermodynamic equilibrium;
for fluids with viscosity and heat flow (see for instance \cite{MTW}) 
new contributions to $dS$
(the term in square brackets in (\ref{relate_1})) 
arise,
on account of
$T_{{\hat 1}{\hat 1}} k^{\hat 1} k^{\hat 1}$ and $2 T_{{\hat 0}{\hat 1}} k^{\hat 0} k^{\hat 1}$ terms
in the integral of (\ref{relate_1}) 
(being $T_{{\hat a}{\hat b}}$ the components of the stress-energy tensor
in the local rest frame of the fluid chosen to have
$k^{\hat 2}, k^{\hat 3} = 0$)  
and corresponding to entropy from viscous heating
and heat flow. 

Let now $k^a$ be a smooth null vector field on our surface $B$, 
everywhere orthogonal to $B$. We can take $k^a$ future
directed without loss of generality. 
Consider the family of null geodesics starting at $B$
with initial tangent $k^a$ (and then with affine parametrization
$\lambda$ such that $k^a = (d/d\lambda)^a$ on $B$ and off $B$,
where $k^a$ is extended off $B$ by parallel transport along the null geodesics).
If $\theta$ is the expansion of the family of geodesics, 
suppose that 
$\theta \leq 0$
everywhere on $B$. 
The hypersurface $L$ to which the bound (\ref{GenBoussoBound}) refers
is then that generated by these null geodesics starting at $B$ 
and terminated at some spacelike 2-surface $B^\prime$
before reaching a caustic ($\theta = -\infty$) or singularity
(and terminating in the caustic or exteded as far as possible
otherwise).
The Raychaudhuri equation applied to this null geodesic congruence
reads 
(the twist term is not reported as it is vanishing as guaranteed 
by its evolution equation starting from the vanishing value on 
$B$ ($k^a$ is surface orthogonal on $B$) \cite{Wald}) 

\begin{eqnarray}\label{Raychaudhuri}
\frac{d\theta}{d\lambda} = -\frac{1}{2}\theta^2 -{\hat \sigma}^2
-R_{ab}k^a k^b, 
\end{eqnarray}
where ${\hat \sigma}^2 = {\hat \sigma}_{ab} {\hat \sigma}^{ab} $ is the square 
of the shear tensor as defined in \cite{Wald}.
Now, denoting with $dA$ and $dA^\prime$ the cross-sectional area
elements respectively
on the surfaces $B$ and $B^\prime$ for a given
geodesic, we have

\begin{eqnarray}\label{Evolution}
dA^\prime - dA = 
dA(\lambda_{max}) - dA(0) =
\int_0^{\lambda_{max}} \theta(\lambda) \ dA(\lambda) \ d\lambda,
\end{eqnarray}
where $\lambda_{max}$ is the value of the affine parameter $\lambda$ on
$B^\prime$ (or on the caustic or singularity)
assuming $\lambda = 0$ on $B$
and $dA(\lambda)$ is such that $dA(0) = dA$.
Using (\ref{Raychaudhuri}), equation (\ref{Evolution}) can be rewritten as

\begin{eqnarray}\label{Evolution2}
\nonumber
dA^\prime - dA = 
\int_0^{\lambda_{max}} \left( \theta(0) + \int_0^\lambda \frac{d\theta}{d\lambda} 
(\lambda^\prime) \ d\lambda^\prime \right) \ dA(\lambda) \ d\lambda
\end{eqnarray}
\begin{eqnarray} 
= \int_0^{\lambda_{max}} \left( \theta(0) - \int_0^\lambda 
\left( \frac{\theta^2}{2} +{\hat \sigma}^2 \right) 
\ d\lambda^\prime \right) \ dA(\lambda)  \ d\lambda
- \int_0^{\lambda_{max}} \left( \int_0^\lambda R_{ab} k^a k^b \ d\lambda^\prime \right)
\ dA(\lambda) d\lambda .
\end{eqnarray}

In this expression the first integral is the sum of three always non-positive terms.
Concerning the second integral,
let us consider the local Rindler horizon \cite{Jacobson}
through the point $p$ on surface $B$ where the geodesic starts
in the direction $k^a$.
In the limit in which
the accelerated worldline approaches the horizon,
this accelerated worldline becomes the geodesic we are considering at $p$.
We can repeat this procedure at every point along the geodesic
so that the accelerated worldline
approaches the geodesic at every point, from $p$ to the
corresponding point $p^\prime$ on $B^\prime$, 
being the local Rindler horizon through each point in the direction
of $k^a$.
Applying the prescription (\ref{key}) to each interval
$(\lambda_i, \lambda_i + \eta)$
of an equally spaced partition of the interval (0,$\lambda_{max}$),
we have

\begin{eqnarray}
4 \ dS_i(p) \leq \frac{1}{\pi (q_i \eta) T_i} \ \int_{\lambda_i}^{\lambda_i + \eta} 
\left( \int_{\lambda_i}^{\lambda} R_{ab} k^a k^b 
\ d\lambda^\prime \right)
\ dA(\lambda) d\lambda
\end{eqnarray}
for small $\eta$,
so that

\begin{eqnarray}\label{entropy}
4 \ dS(p) = \sum_{i=0}^{n-1} 4 \ dS_i(p) \leq 
\sum_{i=0}^{n-1} \frac{1}{\pi (q_i \eta) T_i} \ \int_{\lambda_i}^{\lambda_i + \eta} 
\left( \int_{\lambda_i}^{\lambda} R_{ab} k^a k^b 
\ d\lambda^\prime \right)
\ dA(\lambda) d\lambda
\end{eqnarray}
where $n \eta = \lambda_{max}$ and
with $dS(p)$ and $dS_i(p)$ we denote
the total entropy on the hypersurface spanned
by the horizon area element $dA(\lambda)$ with $\lambda$ going
respectively
from 0 to $\lambda_{max}$
and
from $\lambda_i$ to $\lambda_i + \eta$.

As the second integral $dI_2(p)$ in (\ref{Evolution2}) can be written
as
$ dI_2(p) = \sum_{i=0}^{n-1} \int_{\lambda_i}^{\lambda_i + \eta} 
\left( \int_{0}^{\lambda} R_{ab} k^a k^b \ d\lambda^\prime \right)
\ dA(\lambda) d\lambda$
in the limit $\eta \rightarrow 0$,
we are lead to compare
$C_i \equiv  \frac{1}{\pi (q_i \eta) T_i} \ \int_{\lambda_i}^{\lambda_i + \eta} 
\left( \int_{\lambda_i}^{\lambda} R_{ab} k^a k^b 
\ d\lambda^\prime \right)
\ dA(\lambda) d\lambda$
with
$D_i \equiv \int_{\lambda_i}^{\lambda_i + \eta} 
\left( \int_{0}^{\lambda} R_{ab} k^a k^b \ d\lambda^\prime \right)
\ dA(\lambda) d\lambda$.
To this aim the key point is to estimate the factors
$\frac{1}{\pi (q_i \eta) T_i}$.

One expects a limit on how small can be the scale of definability 
of local equilibrium
and hence on the values of the proper lengths $l_i = q_i \eta$.
This limit has to do with the local thermalization of the material medium
and depends on the physics of this latter;
if we consider a gas, for example, it is given by the mean free path of the molecules.
Let us assume the scale of thermalization be such that

\begin{eqnarray}\label{overestimation}
\frac{1}{\pi l_i T_i} \leq 1, \ \forall i. 
\end{eqnarray}
This inequality appears to be physically plausible (even if not provably universal)
and an argument for its plausibility goes as follows.
Uncertainty principle can be written in the form
$\Delta x \Delta p_x \geq \frac{1}{2}$,
being $\Delta x$ the de Broglie size or spatial quantum uncertainty of matter constituents.
The scale on which local thermodynamic equilibrium will be found is expected in general
much larger and in any case never smaller than the size of the constituents,
whose ultimate lower limit is given by the spatial quantum uncertainty above.

We see thus that in order for equation (\ref{key}) to make sense
as referred to a slice of matter at local equilibrium, 
for the case of material media with
a relation between energy and temperature such that
an energy $T/2$ can be assigned to each spatial degree of freedom,
we should expect 
the proper thickness $l$ of the slice
be such that $l T = 2 l \frac{T}{2} \geq 1$
so that $\frac{1}{\pi l T} \leq \frac{1}{\pi} < 1$.
This is verified, as a concrete example, in the case of a gas.
At realistic conditions in fact the spatial quantum uncertainty of the molecules
is much smaller than their intrinsic size as composite objects, 
so that the mean free path 
is much larger than their de Broglie size
(or, put another way, the times between collisions are much larger
than the temporal quantum uncertainty) and inequality (\ref{overestimation}) is by far satisfied.
For a photon gas, an example with non-composite constituents,
the energy density goes like $T^4$ and hence we obtain 
the spatial uncertainty relation $({\Delta x})^4 T^4 > 1$,
so that, if we assume, according to the above, that local equilibrium length scale
is in general much larger and in any case not smaller than the size
of the constituents, inequality (\ref{overestimation}) follows.
For systems for which the relation between energy and temperature is not linear,
the assumed validity of equation (\ref{overestimation}), 
even if supposedly rooted on the uncertainty principle, 
cannot be obtained by a simple rewriting of 
the latter as above.

According to our assumptions
the point is then now that we are authorized to write total entropy
as a sum on slices with constant entropy as in equation (\ref{entropy})
only provided in each slice
relation (\ref{overestimation}) is satisfied. 
Note that
equation (\ref{overestimation}) turns out to be
the formulation for thin slices of equation (1.9) of \cite{FlanaganMarolfWald}
and recalls equation (1.11) of the same reference or equation (3.5) of \cite{BoussoFlanaganMarolf}.
At the end, it represents some local formulation of the original bound $S \leq \pi E D$
(where $D$ is the diameter of the smallest sphere circumscribing the system
and $E$ is energy)
introduced by Bekenstein \cite{Bekenstein}. 

Again from null convergence condition,
we have
\begin{eqnarray}
C_i \ \leq 
\int_{\lambda_i}^{\lambda_i + \eta} 
\left( \int_{\lambda_i}^{\lambda} R_{ab} k^a k^b \ d\lambda^\prime \right)
\ dA(\lambda) d\lambda \ \leq D_i, \ \forall i
\end{eqnarray}
where in the first inequality relation (\ref{overestimation})
has been used.
From (\ref{entropy})
we thus obtain 
$dI_2(p) \geq 4 \ dS(p)$
and then

\begin{eqnarray}
\nonumber
\frac{A - A^\prime}{4} 
= -\frac{1}{4} \int_0^{\lambda_{max}} \int_{B(\lambda)}
\left( \theta(0) + \int_0^\lambda \frac{d\theta}{d\lambda} 
(\lambda^\prime) \ d\lambda^\prime \right) \ dA(\lambda) \ d\lambda \geq
\end{eqnarray}

\begin{eqnarray}\label{final}   
\geq \int_{B} \frac{dI_2(p)}{4}
\geq \int_{B} dS(p) = S(L)
\end{eqnarray}
that is (\ref{GenBoussoBound}). 

Note that given the null hypersurface $L$,
the argument works with every partition with slices
along the null geodesics generating $L$
satisfying (\ref{overestimation})
and for which entropy can be considered constant with $\lambda$
in each slice.
We see that a partition with a large maximum admissible depth of each slice
gives a strong constraint for $S(L)$.
In general equation (\ref{overestimation}) (and equation (\ref{final})) 
will be by far satisfied.
The only way to challenge (\ref{overestimation})
is to have very thin hypersurfaces $L$, with a thickness quantum-mechanically defined,
or, in general, to have $L$ with so large variations of local entropy
that the depths for which matter entropy con be considered constant
are smaller than what required by the bound (\ref{overestimation}).
As said, in these conditions the mere notion of local thermodynamic equilibrium
appears to get into trouble (see also \cite{FlanaganMarolfWald}).
Looking at equations (\ref{prescription}) and (\ref{key}), 
we see that (\ref{overestimation})
amounts also to say that $dS \leq dS^\prime$ whenever a local thermodynamical description
is viable. This suggests that the entropy element
beyond a spacelike 2-surface should be maximal for an observer
for which that surface is a horizon. 

The discussion below (\ref{overestimation}) is roughly an
order-of-magnitude argument suggesting only that for matter at conditions
of local thermodinamic equilibrium, the bound (\ref{overestimation}),
and then the bound (\ref{GenBoussoBound}),
are by far satisfied.
The reason why the term to the right of (\ref{overestimation}) 
has been chosen to be precisely as large as 1 is to obtain the final
inequality (\ref{final}), that is (\ref{GenBoussoBound}),
i.e. a bound that embodies the GSL 
(assuming the validity of the ordinary second law) \cite{FlanaganMarolfWald}.
Our argument suggests that to saturate this bound
some quantum effects, coming from the system size
or from the strength of the variations of local quantities,
challenging the concept of
local thermodynamic equilibrium should be unavoidable.

The argument presented here,
instead of only a proof of the generalized Bousso bound
(starting from some sufficient conditions), 
appears rather to be
a determination of the conditions on which the bound rests.
What we find in fact is not only that if equation (\ref{overestimation})
holds the bound cannot be violated, 
but also conversely that if equation (\ref{overestimation}) 
can be violated 
in the sense that
local thermodynamic equilibrium is allowed to be defined
at arbitrarily short scales,
necessarily it should then be possible to also violate
the generalized Bousso bound.
If in fact arbitrarily thin
slices can consistently be chosen,
considering without loss of generality a material medium with vanishing chemical potential,
from (\ref{key}) or (\ref{overestimation}) (which turn out to be equalities
in this case) in such slices the bound is violated.

From the uncertainty principle, 
inequality (\ref{overestimation}) can be expected to accompany
any local equilibrium configuration
and can be directly proven at least for a large class of systems, 
as we have seen.
If this is the case, given a classical spacetime with Einstein equation
the generalized Bousso bound would follow
from the mere introduction of quantum mechanics.
In this sense the same physical premise, the quantum principle, 
would cause both the generalized Bousso bound to be satisfied
and black holes, or any local Rindler horizon, to have a temperature.

I thank
the anonymous referee
and R. Bousso for comments and remarks 
on an earlier version of this manuscript,
clarifying
various points on the path 
to the reported results.

\end{document}